\pdfoutput=1
\documentclass{pasj01}
\usepackage{url}
\usepackage{graphicx}
\usepackage{graphicx,todonotes}

\newcommand{\project}[1]{\textsl{#1}}
\newcommand{\secname}{Section}
\newcommand{\figname}{Figure}
\newcommand{\tabname}{Table}

\newcommand{\ngcobj}{NGC~5750}
\newcommand{\sdssobj}{SDSS~J1449-0042}

\newcommand{\sersic}{S\'{e}rsic}
\newcommand{\survey}{HSC-SSP}
\newcommand{\objname}{Sumo Puff}
\newcommand{\resp}[1]{#1}

\newcommand{\sbunits}{mag~arcsec$^{-2}$}

\begin{document}
\SetRunningHead{Greco et al.}{Sumo Puff}

\title{\objname: Tidal Debris or Disturbed Ultra-Diffuse Galaxy?}
\author{Johnny~P.~Greco\altaffilmark{\pu, *},
        Jenny~E.~Greene\altaffilmark{\pu},
        Adrian M.~Price-Whelan\altaffilmark{\pu},
        Alexie Leauthaud\altaffilmark{\santa},
        Song Huang\altaffilmark{\santa},
        Andy D.~Goulding\altaffilmark{\pu},
        Michael A.~Strauss\altaffilmark{\pu},
        Yutaka Komiyama\altaffilmark{\naoj,\sokendai},
        Robert H.~Lupton\altaffilmark{\pu},
        Satoshi Miyazaki\altaffilmark{\naoj,\sokendai},
        Masahiro Takada\altaffilmark{\kavli},
        Masayuki Tanaka\altaffilmark{\naoj},
        Tomonori Usuda\altaffilmark{\naoj,\sokendai}}
        
% Affiliations
\newcommand{\pu}{1}
\altaffiltext{\pu}{Department of Astrophysical Sciences, Princeton University, Princeton, NJ 08544, USA}
\newcommand{\santa}{2}
\altaffiltext{\santa}{Department of Astronomy and Astrophysics, University of California, Santa Cruz, 1156 High Street, Santa Cruz, CA 95064 USA}
\newcommand{\naoj}{3}
\altaffiltext{\naoj}{National Astronomical Observatory of Japan, 2-21-1 Osawa, Mitaka, Tokyo 181-8588, Japan}
\newcommand{\sokendai}{4}
\altaffiltext{\sokendai}{Department of Astronomy, School of Science, Graduate University for Advanced Studies
(SOKENDAI), 2-21-1, Osawa, Mitaka, Tokyo 181-8588, Japan}
\newcommand{\kavli}{5}
\altaffiltext{\kavli}{Kavli Institute for the Physics and Mathematics of the Universe (Kavli IPMU, WPI), University
of Tokyo, Chiba 277-8582, Japan}

\altaffiltext{*}{Corresponding Author}
\email{jgreco@astro.princeton.edu}

\KeyWords{keywords}
\maketitle

\begin{abstract}
We report the discovery of a diffuse stellar cloud with an angular extent $\gtrsim30\arcsec$, which we term ``\objname'', \resp{in data from the Hyper Suprime-Cam Subaru Strategic Program (HSC-SSP)}. While we do not have a redshift for this object, it is in close angular proximity to a post-merger galaxy at redshift $z=0.0431$ and is projected within a few virial radii (assuming similar redshifts) of two other ${\sim}L_\star$ galaxies, which we use to bracket a potential redshift range of $0.0055 < z < 0.0431$. The object's light distribution is flat, as characterized by a low \sersic\ index ($n\sim0.3$). It has a low central $g$-band surface brightness of ${\sim}26.4$~\sbunits, large effective radius of ${\sim}13\arcsec$ (${\sim}11$~kpc at $z=0.0431$ and ${\sim}1.5$~kpc at $z=0.0055$), and an elongated morphology ($b/a\sim0.4$). Its red color ($g-i\sim1$) is consistent with a passively evolving stellar population and similar to the nearby post-merger galaxy\resp{, and we may see tidal material connecting Sumo Puff with this galaxy.} We offer two possible interpretations for the nature of this object: (1) it is an extreme, galaxy-size tidal feature associated with a recent merger event, or (2) it is a foreground dwarf galaxy with properties consistent with a quenched, disturbed ultra-diffuse galaxy. We present a qualitative comparison with simulations that demonstrates the feasibility of forming a structure similar to this object in a merger event. Follow-up spectroscopy \resp{and/or deeper imaging to confirm the presence of the bridge of tidal material} will be necessary to reveal the true nature of this object.
\end{abstract}

\section{Introduction}\label{sec:intro}

Ongoing surveys are now revealing the low-surface-brightness universe as never before. Breakthroughs with observing strategies and data reduction (e.g., \cite{Dalcanton:1997aa, Blanton:2011aa, Ferrarese:2012aa, Duc:2015aa, Fliri:2016aa, Trujillo:2016aa}) and small robotic telescopes \citep{Abraham:2014aa, Javanmardi:2016aa} have allowed us to explore diffuse structures with surface brightnesses $\mu > 25$~\sbunits\ routinely, providing new insights into the full galaxy population \citep{Disney:1976aa}, the halos of massive galaxies \citep{Merritt:2016ab, Mihos:2017aa, Huang:inprep}, and the hierarchical build-up of galaxies \citep{van-Dokkum:2005aa, Tal:2009aa}. Our ongoing Hyper Suprime-Cam Subaru Strategic Program \resp{(\survey; \cite{Aihara:2017ab}), which recently had its first public data release \citep{Aihara:2017aa}}, is imaging the sky with an unprecedented combination of depth and area ($i\sim26$~AB~mag over 1400~deg$^2$ upon completion). This data set will have enormous potential to reveal populations of low-surface-brightness phenomena across a diverse range of environments, opening a new window into the ultra-diffuse universe.

\begin{figure*}[t]
\begin{center}
\includegraphics[width=\textwidth]{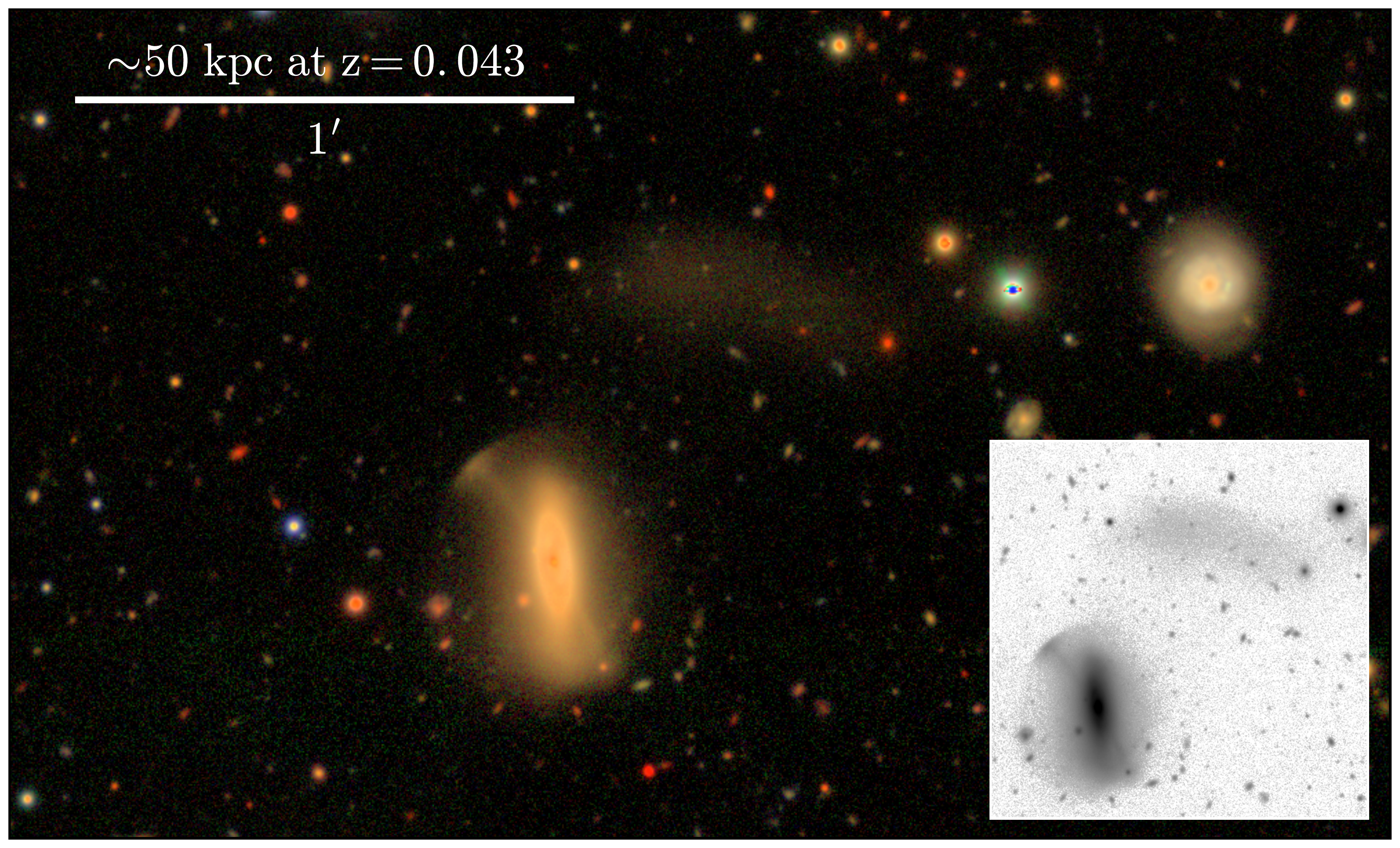}
\end{center}
\caption{\survey\ $gri$-composite image \citep{Lupton:2004aa} of \objname\ and the post-merger galaxy \sdssobj. The inset shows the $i$-band image with a logarithmic stretch. Note the very sharp caustic feature associated with \sdssobj, which is due to a turnaround point in stellar orbits that manifests as a caustic in  surface brightness. If \objname\ is at the redshift of \sdssobj\ (z=0.0431), then it is $>26$~kpc in diameter. North is up, and east is to the left.}
\label{fig:image}
\end{figure*}

Fainter surface brightness limits invariably result in the discovery of new galaxies. This fact was recently demonstrated with the discovery of ${\sim}1000$ so-called ultra-diffuse galaxies in the Coma cluster \citep{van-Dokkum:2015aa, Koda:2015aa}, which are characterized by central surface brightnesses $\mu_0(g)>24$~\sbunits\ and effective radii $r_\mathrm{eff}>1.5$~kpc. Although the existence of such faint and extended galaxies has been known for decades (e.g., \cite{Sandage:1984aa, Impey:1988aa, Bothun:1991aa, Dalcanton:1997aa, McConnachie:2008aa}), such an abundant cluster population was not expected, and this has renewed interest in studying galaxies in the ultra-low-surface-brightness regime. Large-scale, systematic observations optimized at low surface brightnesses will be necessary to uncover the nature of ultra-diffuse galaxies; for example, their number density as a function of environment may reveal what fraction are ``failed'' $L_\star$ galaxies \citep{van-Dokkum:2015aa}, with massive (${\sim}10^{12}~M_\odot$) dark matter halos like Dragonfly 44 \citep{van-Dokkum:2016aa}, versus what fraction represent the high-spin tail of the dwarf galaxy population \citep{Dalcanton:1997ab, Amorisco:2016aa}. Another interesting possibility is that some ultra-diffuse galaxies are ``normal'' dwarf galaxies that are being disrupted by more massive hosts  \citep{Crnojevic:2016aa, Toloba:2016ab, Merritt:2016aa}, similar to dwarf spheroidal galaxies in the Local Group \citep{Collins:2013aa}.

In the $\Lambda$CDM cosmological framework, low-surface-brightness substructure and tidal debris are also the inevitable detritus of the hierarchical build-up of stellar halos. At surface brightnesses fainter than ${\sim}27$--30~\sbunits, massive galaxies are predicted to be engulfed in a rich network of tidal streams, shells, and tails \citep{Bullock:2005aa, Johnston:2008aa, Cooper:2010aa}. Low-surface-brightness observations have confirmed this expectation around nearby elliptical galaxies \citep{van-Dokkum:2005aa, Tal:2009aa, Duc:2015aa}, spiral galaxies in the local volume \citep{Martinez-Delgado:2010aa}, and the high-density cluster environment \citep{Mihos:2017aa}. Such observations contain a wealth of information about the accretion history of massive galaxies. In addition, large samples of the low-surface-brightness substructure and tidal features around nearby galaxies (e.g., \cite{Atkinson:2013aa}) have the potential to constrain the orbital parameter distributions of merging satellites, which would have important implications for the formation of structure in the universe \citep{Hendel:2015aa}.

As we push to unbiased wide area surveys, such as \survey, natural confusion arises between these two sources of low-surface-brightness material (diffuse galaxies and tidal debris). Namely, they sometimes have very similar photometric properties and, when data are limited, can be indistinguishable from one another. While tidal debris is in general clearly connected to a primary system by low-surface-brightness streams and tails, tidal interactions are also capable of producing relatively isolated, giant stellar clouds (e.g., \cite{Johnston:2008aa, Martinez-Delgado:2010aa}). Diffuse-galaxy searches will detect such features as individual galaxies when in fact they are the debris of galaxy interactions. 

In this paper, we report the discovery of an object that falls into this ambiguous category. The object was discovered as part of our ongoing hunt for ultra-diffuse galaxies in data from the \survey. In \secname~\ref{sec:data}, we discuss the data and identification of the object. We present the object's structural, photometric, and environmental properties in \secname~\ref{sec:props}. In \secname~\ref{sec:what-is-it}, we present two possible interpretations of the nature of the object, and we conclude with discussion in \secname~\ref{sec:discussion}.  Throughout this work, we assume standard $\Lambda$CDM cosmological parameters: $\Omega_\mathrm{m}=0.3$, $\Omega_\Lambda=0.7$, and $H_0 = 70~\mathrm{km\ s^{-1}\ Mpc^{-1}}$.

\begin{figure*}[t]
\begin{center}
\includegraphics[width=0.95\textwidth]{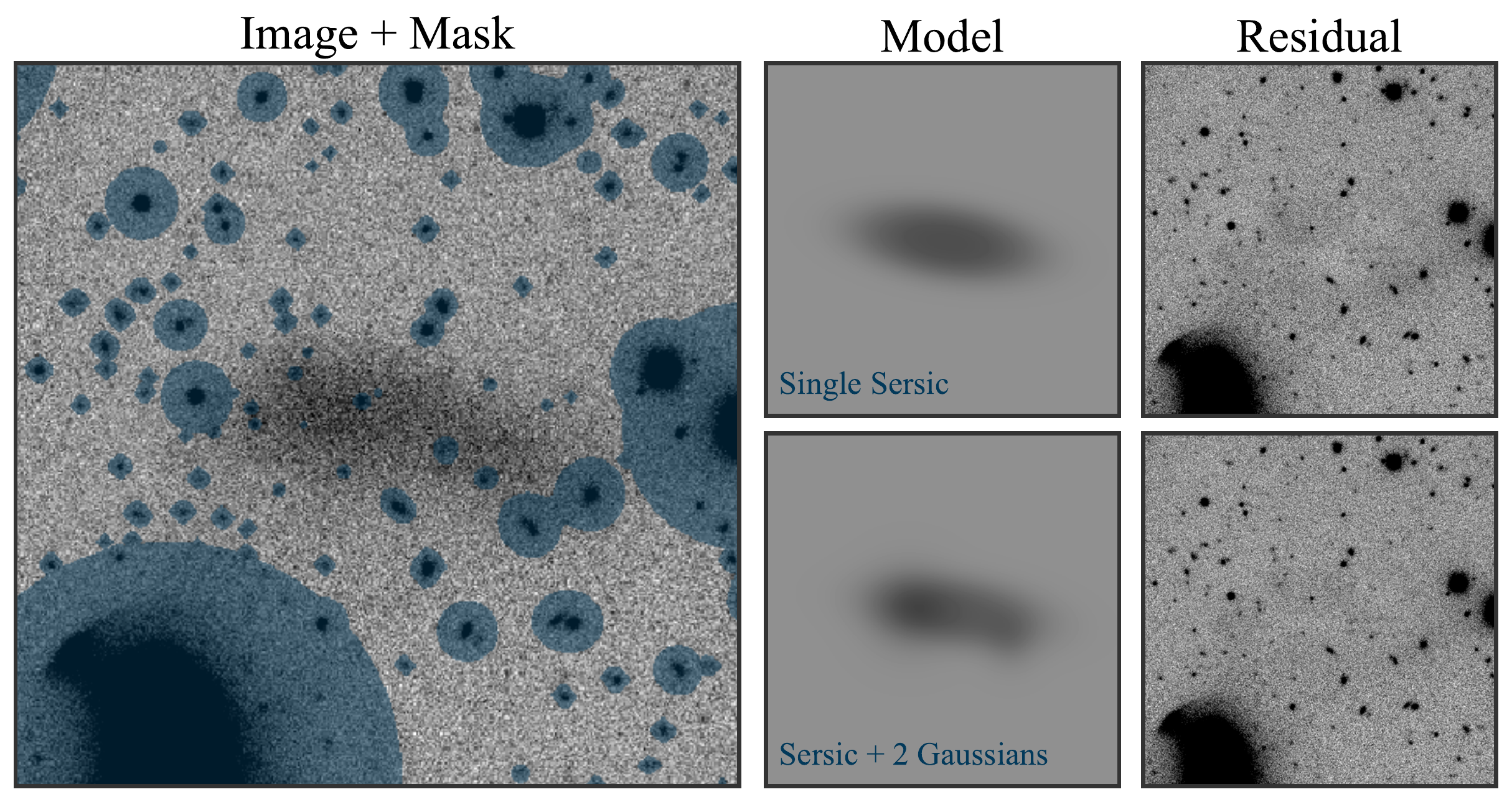}
\end{center}
\caption{{\it Left:} The $r$-band image with an overlay of the object mask. We use the same mask for all bands. {\it Right:} The best-fit single-component (top row) and three-component (bottom row) models and the associated residuals. We summarize the model parameters in \tabname~\ref{tab:props}. The $g$- and $i$-band fits produce qualitatively similar results. The dimensions of each cutout are ${\sim}70\arcsec\!\times70\arcsec$.} 
\label{fig:model-fits}
\end{figure*}

\section{Data and Identification}\label{sec:data}

The data were taken with the Hyper Suprime-Cam (HSC; \cite{Miyazaki:2012aa}) on the 8.2-meter Subaru Telescope as part of the wide layer of the \survey, \resp{an ambitious 300-night, wide-field multi-filter} imaging survey \citep{Aihara:2017ab}. Utilizing HSC's large field of view (1.5~deg in diameter) and the large telescope aperture, this survey will achieve $i\sim 26$~AB~mag ($5\sigma$ point-source depth) over $1400$ deg$^2$ upon completion. Standard data reductions were performed using \project{hscPipe}, a modified version of the Large Synoptic Survey Telescope (LSST) software stack \citep{Ivezic:2008aa, Axelrod:2010aa, Juric:2015aa}. For details about \project{hscPipe} and the \survey\ data reductions, see \citet{Bosch:2017aa} and \citet{Aihara:2017aa}.

Taking advantage of the unprecedented combination of depth and wide-area coverage afforded by \survey, we are using this data set to carry out a systematic search for ultra-diffuse galaxies outside of cluster environments \citep{Greco:2017aa}. Currently, \project{hscPipe} is not optimized to find low-surface-brightness objects. With central surface brightnesses ${\sim}3$--$5$ magnitudes fainter than the night sky, the light from such sources is easily dominated by background/foreground objects. As a result, the \project{hscPipe} photometric deblender shreds them into many smaller constituent parts, a problem that is well known to exist in the Sloan Digital Sky Survey (SDSS; \cite{York:2000aa}) but is much more pronounced at the depth of \survey. Therefore, we are developing custom tools based on the LSST software stack to perform our diffuse-galaxy search. 

During the development of our software, we discovered a giant stellar cloud, ``\objname'', whose origin and physical nature is currently ambiguous. In \figname~\ref{fig:image}, we show its $gri$-composite image. This source was selected by our automated pipeline due to its large angular size and low ``central'' surface brightness. The details of our pipeline will be given in a paper devoted to our primary search. 

We note that \objname\ is clearly visible in all five bands covered by \survey\ ($grizy$) and is detected in individual exposures, which are taken at a range of dither positions. \resp{For $g$ and $r$, every point within the wide layer footprint is observed with 4 exposures per band arranged on a hexagonal grid with the boresight offset by 36$^\prime$; for $izy$, a similar pattern is used except with 6 exposures per band (see \citet{Aihara:2017ab} for details of \survey's observing strategy). Thus, we are confident that \objname\ is a real astrophysical source as opposed to an optical artifact such as a ghost or scattered light from a bright star.} 

\resp{At the depths of \survey, optical scattered light from galactic cirrus can mimic unresolved stellar systems (e.g., \cite{Duc:2015aa, Miville-Deschenes:2016aa}). While the structure of \objname\ appears qualitatively different from the wispy structure characteristic of galactic cirrus, it is nevertheless important to rule out this possibility. If \objname\ is a Milky Way cirrus cloud, then it will likely have a mid/far-infrared counterpart with thermal dust emission. We searched for such a counterpart in the galactic dust map of \citet{Meisner:2014aa}, which is based on a reprocessing of the Wide-field Infrared Survey Explorer 12~$\mu$m imaging data set. We find no potential infrared counterpart. The region of sky ${\sim}2\degree$ around \objname\ has relatively little 12~$\mu$m emission. At \objname's location, there are no visible structures, and the 12~$\mu$m signal is well within $1\sigma$ of the background level.}

\section{Observed Properties of \objname}\label{sec:props}

In \survey\ images, \objname\ appears as a diffuse cloud with unresolved stellar populations, an enormous apparent size ($\gtrsim30\arcsec$ in diameter), and an elongated, peanut-like morphology (see \figname~\ref{fig:image}). Furthermore, it is in close angular proximity to a visually striking post-merger galaxy (SDSS~J144916.46-004240.5; \sdssobj\ hereafter) with a similar red color, which hosts a massive tidal stream with a sharp turnaround point in stellar orbits that manifests as a caustic in surface brightness \citep{Tremaine:1999aa}. It is tempting to assume that \objname\ is tidal debris associated with \sdssobj. In this work, we consider this possibility along with the alternative hypothesis that it is a foreground dwarf galaxy with properties that are consistent with a disturbed ultra-diffuse galaxy.

\subsection{Structure and Photometry}\label{sec:structure}

\begin{figure*}[t] 
\begin{center}
 \includegraphics[width=\textwidth]{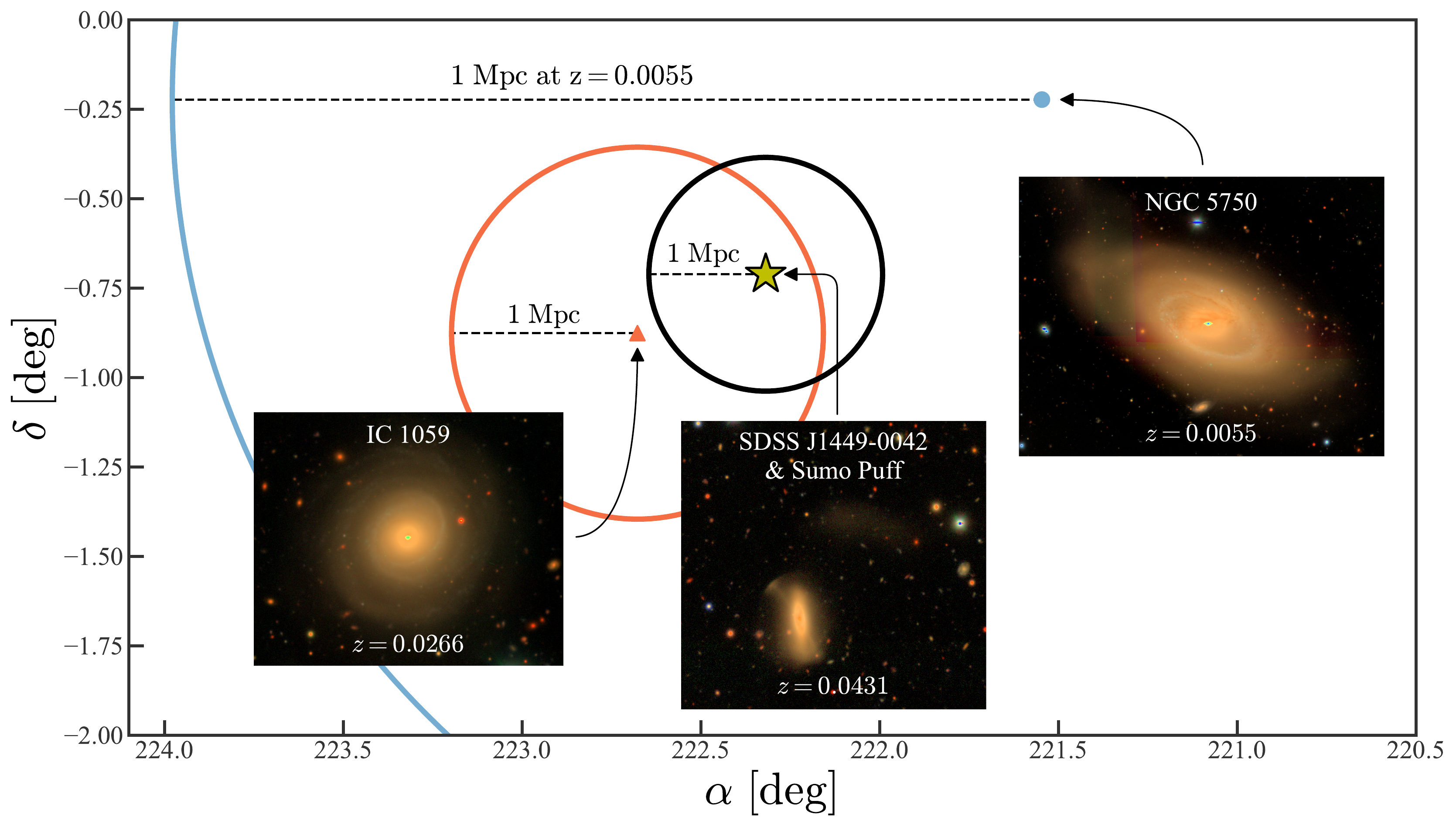}
 \end{center}
 \caption{Positions of \objname\ and three potential host galaxies, which are each the nearest ${\sim}L_\star$ galaxy to \objname\ (assuming similar redshifts) in the redshift ranges we selected from the NSA galaxy catalog (see \secname~\ref{sec:env} for details). For reference, we also show HSC cutout $gri$-composite images of each host candidate and circles representing their associated 1~Mpc search region. This search radius corresponds to ${\sim}3$--$4$ virial radii for the Milky Way and encompasses the region within which we expect to find quenched dwarf galaxy satellites around massive hosts. Due to \ngcobj's very low redshift, its circle extends beyond the limits of the figure. Note that the relative  sizes of the cutout images are not to scale.}
\label{fig:nearby-gals}
\end{figure*}

\begin{table}[htbp]
\begin{center}
\begin{tabular}{cc}
\hline
Parameter & Value\\
\hline
$\alpha_{2000}$  & $222.31310\degree$\\
$\delta_{2000}$  & $-0.70188094\degree$\\
$r_\mathrm{eff}$ & 13.2\arcsec \\
\sersic\ $n$     & 0.32 \\
$b/a$            & 0.38 \\
$m_g$            & 20.0 \\
$g - i$          & 1.0 \\
$g - r$          & 0.68 \\
$\mu_0(g)$       & 26.4 \sbunits \\
$\mu_\mathrm{eff}(g)$ & 26.8 \sbunits\\
\hline\hline
\end{tabular}
\end{center}
%\vspace{0.2cm}
\caption{\objname\ properties. Photometric and structural properties assume a single-component \sersic\ model. The total magnitudes from our independent multi-component fits are within ${\sim}0.02$~mag of those presented in this table.}
\label{tab:props}
\end{table}

To characterize the structural and photometric properties of \objname, we run \project{imfit}\footnote{\url{http://www.mpe.mpg.de/~erwin/code/imfit/}} \citep{Erwin:2015aa} on the sky-subtracted $g$-, $r$-, and $i$-band images. When performing the fit, we use the noise images, object masks, and point-spread function (PSF) measurements generated by the standard \survey\ photometric pipeline. We mask additional objects by hand, which fall within the source footprint and are not covered by the \survey\ object masks. For simplicity, we model the surface brightness distribution as a two-dimensional, PSF-convolved \sersic\ function \citep{Sersic:1968aa}. We first fit each band separately, allowing all parameters to vary; the resulting shape parameters differed by only a few percent across the bands. We adopt their mean values, which are summarized in \tabname~\ref{tab:props}. To measure the color, we perform the fits again on each band with the shape parameters fixed to their mean values, allowing only the amplitude to vary. We find the light distribution to be very flat, characterized by a low \sersic\ index ($n\sim0.3$). However, it is not smooth on large scales. The residual image indicates a lumpy morphology that is not well-captured by a single-component fit. 

Therefore, we also measure the total flux using a three-component model composed of a \sersic\ function and two elliptical Gaussians. In each band, we convolve this multi-component model with the PSF before comparing with the imaging data. In \figname~\ref{fig:model-fits}, we show the $r$-band image with an overlay of the object mask, the best-fit single- and multi-component models, and the associated residuals. The $g$- and $i$-band fits show qualitatively similar results. The total magnitudes of the single- and multi-component models are consistent at the ${\sim}0.02$~mag level. We adopt the single-component \sersic\ model parameters, which are summarized in \tabname~\ref{tab:props}. 

\subsection{Environment}\label{sec:env}

Given \objname's morphology and apparent proximity to the post-merger \sdssobj, it is tempting to associate the two objects; however, an equally viable possibility given the available data is that it is a foreground dwarf galaxy, whose alignment with  \sdssobj\ is the result of chance. Assuming \objname\ is a foreground dwarf galaxy, its red color---which is consistent with a passively evolving stellar population---suggests it has at least one luminous companion; quenched dwarf galaxies are rarely ($<\!0.06\%$) found in isolation, almost exclusively being within 2 virial radii of a massive galaxy \citep{Geha:2012aa}.  Thus, we bracket a potential distance range to \objname\  using the redshifts of its nearest neighbors on the sky. 

We carry out a search for potential massive hosts using the NASA-Sloan Atlas\footnote{\url{http://nsatlas.org}} (NSA), which includes virtually all galaxies with known redshifts out to $z<0.055$ within the coverage of SDSS DR8 \citep{Aihara:2011aa}. We consider galaxies with redshifts $0.002 < z < 0.055$ and stellar masses $\log_{10}(M_\star/M_\odot) > 10$. The low-$z$ cut removes optical artifacts, bright stars, and galaxies with negative redshifts; and the high-$z$ cut is set by the NSA catalog and extends beyond the redshift of the post-merger galaxy \sdssobj\ ($z=0.0431$) seen in \figname~\ref{fig:image}. The stellar masses in the NSA are calculated using \project{kcorrect} \citep{Blanton:2007aa}, which assumes the initial mass function of \citet{Chabrier:2003aa}. We then select any galaxy with a comoving separation less than 1~Mpc from \objname\  {\it assuming} that they are at the same redshift. For the Milky Way, this separation corresponds to  ${\sim}3$--4 virial radii (\cite{Bland-Hawthorn:2016aa}, and references therein), which encompasses the region within which the vast majority of quenched dwarf galaxies are expected to reside near a massive host \citep{Geha:2012aa}. 

The above search produces 7 galaxies that fall roughly into 3 redshift bins: $z \sim 0.006$, 0.03, and 0.04. Of these 7 galaxies, we select the nearest to \objname\ (assuming similar redshifts) in each redshift bin as potential hosts, yielding 3 host galaxy candidates. In \figname~\ref{fig:nearby-gals}, we show the position of \objname\ along with the potential hosts, all of which have stellar masses comparable to $L_\star$ galaxies. For reference, we also show HSC cutout images of each host candidate and circles representing their associated 1~Mpc search region. Without knowing \objname's distance, it is possible that it is a member of any of these systems. The lowest redshift host candidate is \ngcobj\ at $z=0.0055$, and the highest redshift host candidate is the post-merger \sdssobj\ at $z=0.0431$. We, therefore, assume \objname's potential redshift range to be $0.0055 < z < 0.0431$. 

If \objname\ is in fact tidal debris associated with \sdssobj, then it is at $z=0.0431$. On the other hand, if it is a satellite dwarf galaxy, then it may be at any redshift in the above range, with three potential host galaxies shown in \figname~\ref{fig:nearby-gals} --- even at the smallest possible distance in the assumed range (${\sim}24$~Mpc), \objname's physical properties would be consistent with a disturbed ultra-diffuse galaxy. We consider each of these scenarios below.

\section{What is \objname?}\label{sec:what-is-it}

\subsection{Tidal debris associated with \sdssobj}

\begin{figure}[t!]
\begin{center}
\includegraphics[width=8.2cm]{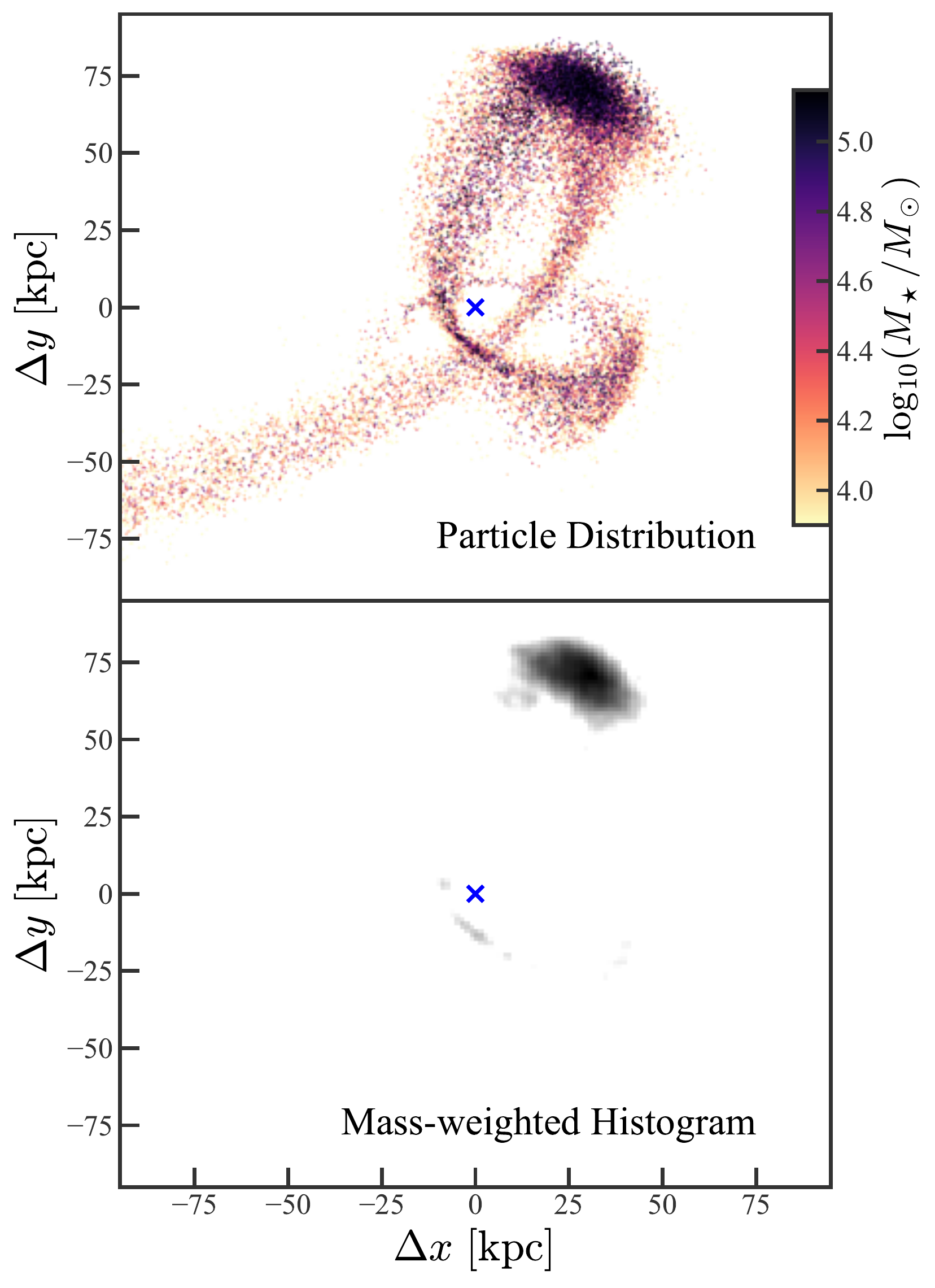}
\end{center}
\caption{{\it Top:} Particle positions of satellite number 11 in halo 15 from the stellar halo models of \citet{Bullock:2005aa}. Particle stellar masses are indicated by color, and the blue cross marks the location of the Milky-Way-like host galaxy. {\it Bottom:} Mass-weighted histogram of the particle distribution smoothed with a Gaussian kernel and shown on an arbitrary logarithmic scale. Assuming a constant mass-to-light ratio, this roughly represents a possible observation of this satellite. Although the scale is arbitrary, it demonstrates that it is possible to create a structure similar to \objname\ in a merger event with all evidence of a tidal stream hidden below an instrument's detection limits and/or behind the main galaxy. Note this figure is meant to provide a qualitative example of a plausible merger geometry and is not an attempt to model the full details of the merger seen in \figname~\ref{fig:image}.}
\label{fig:satellite}
\end{figure}

\begin{figure*}[t!]
\begin{center}
\includegraphics[width=\textwidth]{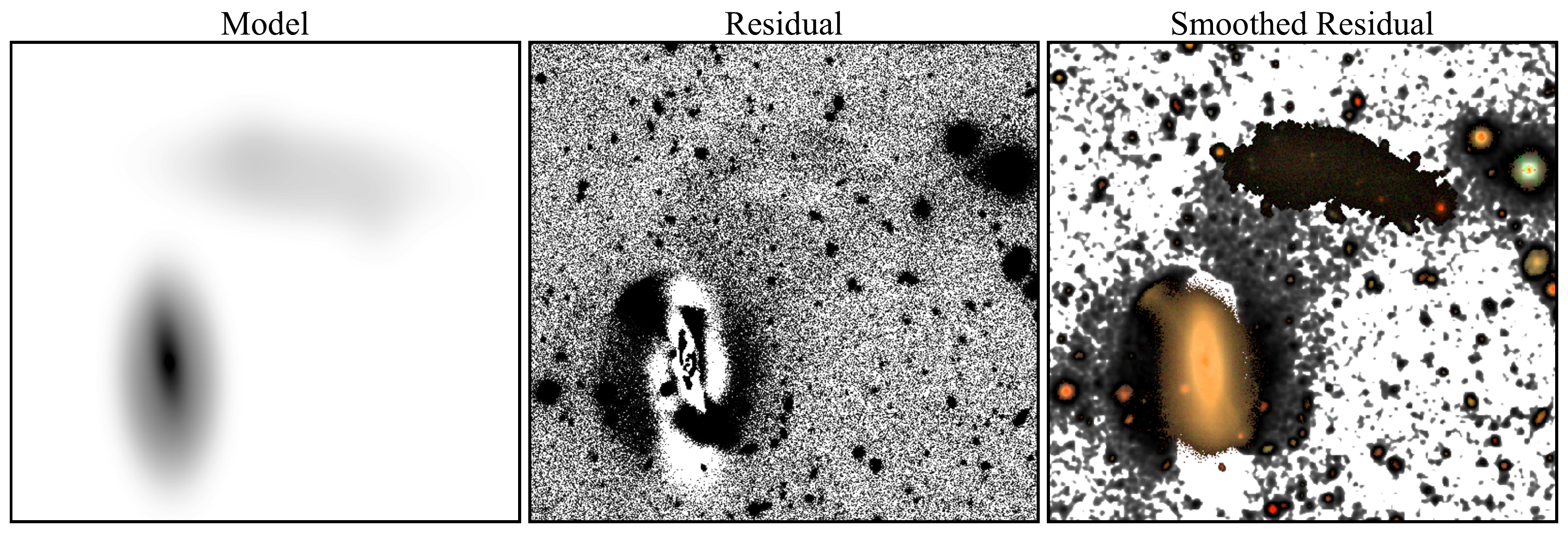}
\end{center}
\caption{
\resp{Tentative evidence of a bridge of low-surface-brightness material connecting \objname\ and \sdssobj. We show a model of both objects (left), the residual image after subtracting this model from the $i$-band image (middle), and the residual image smoothed with a Gaussian kernel to accentuate faint structures (right). We have overlaid sources from the $gri$-composite image on the smoothed residual image to help guide the eye.}
}
\label{fig:bridge}
\end{figure*}

In this scenario, \objname\ is not actually a galaxy at all. Rather, it is tidal debris associated with the recent merger that is apparent in \sdssobj\ at $z=0.0431$. It has a dramatic effective radius of $r_\mathrm{eff}\sim11$~kpc and an $i$-band absolute magnitude of $M_i\sim-17.4$~mag. Using the mass-to-light ratio/color relation derived from \citet{Bell:2003aa}, this corresponds to ${\sim}1.3\times10^9~M_\odot$ of stars. At this redshift, \objname\ is projected within ${\sim}20$--$30$~kpc from the center of \sdssobj\ --- well within its virial radius. If true, this scenario naturally explains \objname's lumpy and elongated ($b/a\sim0.4$) morphology, red color ($g-i\sim1$), and close proximity to \sdssobj. 

It is interesting to speculate on the possible geometry leading to this very extended yet remote tidal feature, which appears to lack any sharp edges in its surface brightness profile. Qualitatively similar giant stellar clouds have been observed in the diffuse stellar halos of nearby spiral galaxies \citep{Martinez-Delgado:2010aa, Martinez-Delgado:2012aa}, the tidal streams of dwarf galaxies \citep{Rich:2012aa, Martinez-Delgado:2015aa}, and in simulations of the formation of stellar halos around Milky-Way-type galaxies \citep{Bullock:2005aa, Johnston:2008aa}. For example, the left panel of \figname~17 of \citet{Johnston:2008aa} shows a large plume with some resemblance to \objname; the authors attribute such features to the accretion of a satellite on a radial orbit. However, shell-like structures viewed from most angles have caustic-like features or sharp edges in their surface brightness profiles, unlike \objname.

To investigate the possibility of forming an apparently caustic-free, isolated stellar cloud similar to \objname\ via the accretion of a satellite, we use a stellar halo model (halo~15) from the  \citet{Bullock:2005aa} simulations\footnote{The models are available at \url{http://user.astro.columbia.edu/~kvj/halos/}.} (note that here we are only considering \objname\ and not the pronounced shell and more compact tidal features apparent in the main galaxy). These models use a hybrid $N$-body and semi-analytic approach to study the build-up of stellar halos of Milky-Way-like galaxies through the disruption and accretion of satellite galaxies within the $\Lambda$CDM cosmological context. For full details of the methods used in these simulations, see \citet{Bullock:2005aa}, \citet{Robertson:2005aa}, and \citet{Font:2006aa}. 

For our purposes, we are simply interested in finding an accretion event that, when viewed from a particular angle, creates a structure that is qualitatively similar to \objname. Note that we are not attempting to model the full details of the merger seen in \figname~\ref{fig:image}; rather, our aim is to provide a qualitative example of a possible merger geometry. In the top panel of \figname~\ref{fig:satellite}, we show the particle positions of satellite number 11 projected onto the two-dimensional viewing plane. Particle stellar masses, which are indicated by color, are not uniform due to the particle tagging method used in \citet{Bullock:2005aa}. This satellite's initial stellar mass was $\log_{10}(M_\star/M_\odot)\sim8.8$. The blue cross marks the location of the host galaxy. In the bottom panel, we show a mass-weighted, two-dimensional histogram of the particle positions on an arbitrary logarithmic scale. Assuming a constant mass-to-light ratio (not a great assumption, but likely good enough for the point being made here), mass is proportional to flux, and this histogram roughly represents what one might see in an observation of this satellite. Although the scaling is arbitrary, it nevertheless demonstrates that there is a scale (set by the detection limits of the observation) at which a caustic-free, isolated stellar cloud will be visible with essentially all evidence of a tidal stream hidden below the instrument's detection limits and/or behind the main galaxy. 

We note that the combination of \objname\ and the sharp shell-like feature in the upper-left region of the post-merger \sdssobj\ (see \figname~\ref{fig:image}) is still quite puzzling, since both structures appear on the same northern side of the main galaxy. \resp{This may suggest that they are due to a single merger event, but it is also possible} that they were created during independent merger events. However, as \figname~\ref{fig:satellite} demonstrates, projection effects combined with surface brightness limits can produce non-intuitive geometries, which are difficult to understand without the aid of more tailored simulations. \resp{We, therefore, remain agnostic about whether a single or multiple merger events are responsible for the morphology of this system.} 

\resp{
Close inspection of the images shown in \figname s~\ref{fig:image} and \ref{fig:model-fits} reveals a possible bridge of low-surface-brightness material connecting \objname\ and \sdssobj. If this feature is real, it would strongly support the tidal scenario. To ensure that we are not simply observing the overlapping isophotes from these two sources, we subtract a model of both objects from the $i$-band image to reveal faint structures that are hidden below the dominant components of the light profiles. For \objname, we use the multi-component model presented in \secname~\ref{sec:structure}. For \sdssobj, we again use \project{imfit} to fit the galaxy's surface brightness profile with a multi-component model consisting of two \sersic\ functions and a Gaussian function at the center of the galaxy. When performing the fits, we mask sources (including the sharp shell-like feature on the upper-left of the galaxy) that are not associated with the galaxy's smooth light profile.
}

\resp{
The results are shown in \figname~\ref{fig:bridge}. In the left panel, we show the combined model for \objname\ and \sdssobj. In the middle panel, we show the residuals after subtracting this model from the $i$-band image. Within the main galaxy, we see fine structure that is associated with its merger history and morphological properties that are not captured by our simple model. In addition, there is a hint of a bridge between the two objects. To emphasize this feature, we smooth the residual image with a Gaussian kernel with a full width at half maximum of 0.8\arcsec\ (slightly larger than the PSF). We show the smoothed residual image in the right panel with sources from the $gri$-composite image overlaid for reference. The smoothing accentuates the low-surface-brightness bridge, which provides tentative evidence in support of the tidal scenario for \objname. Deeper images could confirm the presence of this bridge material and would argue strongly that Sumo Puff is tidal.
} 

\resp{
Considering the evidence presented in this section, we favor a scenario in which \objname\ is tidal debris from a satellite that is being accreted by \sdssobj. This accretion event is not necessarily associated with the more compact tidal features that are clearly present in \sdssobj.   
}

\subsection{Disturbed ultra-diffuse galaxy}\label{foreground}

In this alternative scenario, \objname\ is a dwarf galaxy, whose red color suggests it is a satellite of a massive companion \citep{Geha:2012aa}. As described in \secname~\ref{sec:env}, we have selected three potential host galaxies that fall within the redshift range $0.0055 < z < 0.0431$, corresponding to a comoving distance range ${\sim}24$--180~Mpc. At the low end of this distance range, \objname\ would have an effective radius $r_\mathrm{eff}\sim1.5$~kpc and absolute $i$-band magnitude $M_i\sim-12.9$~mag, which is consistent with the size-luminosity relation for dwarf/ultra-diffuse galaxies (e.g., \cite{Munoz:2015aa}). Using the same mass-to-light ratio/color relation derived from \citet{Bell:2003aa} above, this implies a stellar mass $M_\star\sim2.0\times10^7~M_\odot$. With a central $g$-band surface brightness of ${\sim}26.4$~\sbunits, these physical properties would classify \objname\ as an ultra-diffuse galaxy. In addition, \objname's lumpy and elongated morphology would suggest it is being tidally disrupted, possibly similar to objects recently discovered around the massive ellipticals NGC~5485 and NGC~5473 at a similar distance from the Milky Way \citep{Merritt:2016aa}. There are also similarities with the only known ultra-diffuse galaxy in the Local Group, And~XIX \citep{McConnachie:2008aa}, which is ${\sim}187$~kpc from M31 and is known to be disrupting \citep{Collins:2013aa}.

Our diffuse-galaxy search covered ${\sim}$100~deg$^2$ at the time of \objname's discovery. Given this small search area and the little volume that is probed out to $z = 0.0055$ (the lowest redshift in our assumed range), how many \objname-like galaxies with $z < 0.0055$ do we expect in our sample? Assuming the luminosity function of \citet{Blanton:2005aa}, which explicitly corrects for low-surface-brightness galaxies, we expect our sample to contain ${\sim}0.5$ galaxies with absolute magnitude and surface brightness comparable to \objname\ (assuming $z=0.0055$). However, this expectation is based on a large extrapolation, since our object is nearly a magnitude fainter than the faintest luminosities in the \citet{Blanton:2005aa} study, and its surface brightness is well below the limits of their survey. With the imaging of \survey, we are beginning to push down into this relatively unexplored region of parameter space, which will enable the extension of the luminosity function of field galaxies down to such low luminosities. 

In this dwarf-galaxy scenario, \objname's physical properties become increasingly extreme with increasing redshift. At the high end of the assumed redshift range ($z=0.0431$), its properties would be the same as given in the above tidal scenario: $r_\mathrm{eff}\sim11$~kpc and $M_i\sim-17.4$~mag. These properties would make it somewhat of an outlier in the size-luminosity relation for galaxies, with a size comparable to the most extreme ultra-diffuse galaxies known, VLSB-A and VLSB-D in the Virgo cluster \citep{Mihos:2015aa, Mihos:2017aa}, which are also thought to be undergoing tidal disruption. However, assuming a redshift of $z=0.0431$, \objname's total luminosity would be ${\sim}2$ magnitudes brighter than these objects.

\section{Discussion and Conclusions}\label{sec:discussion}

We have reported the discovery of a potentially giant, low-surface-brightness stellar cloud (\objname), which is in close proximity on the sky to the post-merger galaxy SDSS~J144916.46-004240.5 (\sdssobj; \figname~\ref{fig:image}). We discovered this remarkable object during the development of our automated pipeline to detect diffuse galaxies in the wide layer of \survey. We offer two possible interpretations for the nature of \objname: (1) it is an extreme, galaxy-size tidal feature \resp{produced by a recent merger event with} \sdssobj, or (2) it is a disturbed ultra-diffuse galaxy in the redshift range $0.0055 < z < 0.0431$, with properties that become increasingly extreme with increasing redshift. Because of \objname's close angular proximity to \sdssobj, its lumpy and elongated morphology (\tabname~\ref{tab:props} and \figname~\ref{fig:model-fits}), qualitative similarities to merger simulations (\figname~\ref{fig:satellite}), \resp{and the potential bridge of low-surface-brightness material connecting \objname\ and \sdssobj\ (\figname~\ref{fig:bridge})}; we prefer \resp{the tidal debris scenario. In this scenario, \objname\ may or may not be the result of the same accretion event that caused the pronounced shell and more compact tidal features apparent in \sdssobj. A distance measurement and/or deeper imaging to confirm the existence of the bridge material will be necessary to reveal the true nature of \objname.}

If our interpretation is correct, we are watching the halo of this galaxy being built. This finding demonstrates the potential of deep {\it and} wide surveys to inform theories of galaxy formation by testing the predictions of hierarchical growth of structure in a $\Lambda$CDM universe. Upon completion, the wide layer of \survey\ will reach depths of  $i\sim26$ over 1400~deg$^2$. Using reduced \survey\ (full survey depth) images, which have not been optimized for low-surface-brightness science, we are currently detecting objects with central surface brightnesses as low as $\mu_0(i)\sim26.5$--$27$~\sbunits\ \citep{Greco:2017aa} and the outskirts of massive ellipticals have been measured down to $\mu(i)\sim28.5$~\sbunits\ \citep{Huang:inprep}. Simulations predict that the halos of massive galaxies are rich with substructure at these surface brightnesses \citep{Johnston:2008aa}. Therefore, the full \survey\ data set will have the potential to provide a statistical sample of extragalactic tidal streams and \objname-like objects, which can, in principle, reveal properties of the underlying population of satellites that are being accreted to form the halos of massive galaxies. 

\objname\ was discovered as part of our ongoing search for ultra-diffuse galaxies outside of cluster environments. If it is in fact tidal in nature, then it is a \resp{scientifically interesting} contaminant in our search. As the discovery space of these elusive galaxies pushes down to lower-density environments, it will become increasingly important to understand all potential sources of contamination. As \objname\ demonstrates, confusion between tidal debris and disrupting dwarf galaxies may be significant. For example, some of the objects discovered by \citet{Merritt:2016aa} may actually be tidal in origin. On the other hand, if \objname\ is a foreground dwarf galaxy, it is a very nearby example of an ultra-diffuse galaxy with interesting morphological properties.

\section*{Acknowledgments}

We thank Kathryn Johnston, David Hendel, and David Spergel for useful discussions and Jim Gunn and Sebastien Peirani for providing insight into the possible origin of \objname. We also thank Kathryn Johnston for sharing simulation data, which informed this work. J.P.G. is supported by the National Science Foundation Graduate Research Fellowship under Grant No. DGE 1148900. J.E.G. is partially supported by NSF AST-1411642.

The Hyper Suprime-Cam (HSC) collaboration includes the astronomical communities of Japan and Taiwan, and Princeton University. The HSC instrumentation and software were developed by the National Astronomical Observatory of Japan (NAOJ), the Kavli Institute for the Physics and Mathematics of the Universe (Kavli IPMU), the University of Tokyo, the High Energy Accelerator Research Organization (KEK), the Academia Sinica Institute for Astronomy and Astrophysics in Taiwan (ASIAA), and Princeton University. Funding was contributed by the FIRST program from Japanese Cabinet Office, the Ministry of Education, Culture, Sports, Science and Technology (MEXT), the Japan Society for the Promotion of Science (JSPS), Japan Science and Technology Agency (JST), the Toray Science Foundation, NAOJ, Kavli IPMU, KEK, ASIAA, and Princeton University. 

This paper makes use of software developed for the Large Synoptic Survey Telescope. We thank the LSST Project for making their code available as free software at http://dm.lsst.org.

The Pan-STARRS1 Surveys (PS1) have been made possible through contributions of the Institute for Astronomy, the University of Hawaii, the Pan-STARRS Project Office, the Max-Planck Society and its participating institutes, the Max Planck Institute for Astronomy, Heidelberg and the Max Planck Institute for Extraterrestrial Physics, Garching, The Johns Hopkins University, Durham University, the University of Edinburgh, Queen‚Äôs University Belfast, the Harvard-Smithsonian Center for Astrophysics, the Las Cumbres Observatory Global Telescope Network Incorporated, the National Central University of Taiwan, the Space Telescope Science Institute, the National Aeronautics and Space Administration under Grant No. NNX08AR22G issued through the Planetary Science Division of the NASA Science Mission Directorate, the National Science Foundation under Grant No. AST-1238877, the University of Maryland, and Eotvos Lorand University (ELTE) and the Los Alamos National Laboratory.

Based in part on data collected at the Subaru Telescope and retrieved from the HSC data archive system, which is operated by Subaru Telescope and Astronomy Data Center, National Astronomical Observatory of Japan.

This research additionally utilized: \texttt{Astropy} \citep{Astropy-Collaboration:2013aa}, \texttt{IPython} \citep{Perez:2007aa}, \texttt{matplotlib} \citep{Hunter:2007aa}, and \texttt{numpy} \citep{Van-der-Walt:2011aa}.

\bibliographystyle{apj}
\bibliography{puffball-min}
\end{document}